\documentclass[12pt,twoside]{article}
\usepackage{fleqn,espcrc1} 
\usepackage{graphicx,wrapfig,epsfig}
\usepackage[figuresright]{rotating}

\newcommand{\AmS}{{\protect\the\textfont2
  A\kern-.1667em\lower.5ex\hbox{M}\kern-.125emS}}
\begin{document}

\title{Charged particle multiplicity and transverse energy in Au - Au collisions at $\sqrt{s_{_{NN}}}=130$ GeV}
\maketitle
A. Milov for the PHENIX Collaboration\footnote{For the full PHENIX Collaboration author list and acknowledgements see the contribution by W.A.~Zajc (K. Adcox {\it et al.}) in this volume.}
\vspace{2mm}

Weizmann Institute of Science, Israel
\vspace{2mm}


\begin{abstract}
This paper presents the results for the charged-particle multiplicity and transverse energy distributions at mid-rapidity in Au - Au collisions at $\sqrt{s_{_{NN}}}=130$~GeV measured with the PHENIX detector at RHIC. The values of $dN_{ch}/d\eta_{|\eta=0}$ and $dE_{T}/d\eta_{|\eta=0}$, analyzed as a function of centrality, show a consistent steady rise. For the 5\% most central collisions they are $\sim70$\% larger compared to the SPS results for Pb-Pb collisions at $\sqrt{s_{_{NN}}}=17.2$~GeV. The ratio of $\langle E_{T}\rangle/\langle N_{ch}\rangle$ remains constant as a function of centrality at $0.8$~GeV, as also observed at the SPS at CERN and at the AGS at BNL.
\end{abstract}

\section{INTRODUCTION}
Global variables such as $dN_{ch}/d\eta_{|\eta=0}$ and $dE_{T}/d\eta_{|\eta=0}$ are important for the characterization of high-energy nuclear collisions. They constrain theoretical models, help to discriminate among various mechanisms of entropy and particle production and allow to derive information about the initial conditions~\cite{bass-qm99}. During the first run of RHIC at $\sqrt{s_{_{NN}}}=130$~GeV charged particle multiplicity and transverse energy density distributions were measured using the PHENIX detector. The results are analyzed as a function of centrality, defined by the number of nucleons participating in the collision and compared to similar results obtained at the much lower energies of the CERN SPS and the BNL AGS.

\section{THE PHENIX DETECTOR}
A detailed description of the PHENIX detector can be found in~\cite{phenix-detector}. The detector consists of an axial-field magnet with two central arms, the East and the West arms, each one subtending 90$^{o}$ in azimuth ($\phi$) and $\pm 0.35$ units of pseudo-rapidity ($\eta$). Two spectrometers, the North and South spectrometers, at forward angles ($10^{o}<\theta<35^{o}$) around the beam axis serve to identify and track muons.

\begin{wrapfigure}{l}{10.3cm}
\vspace{-7mm}
\epsfig{file=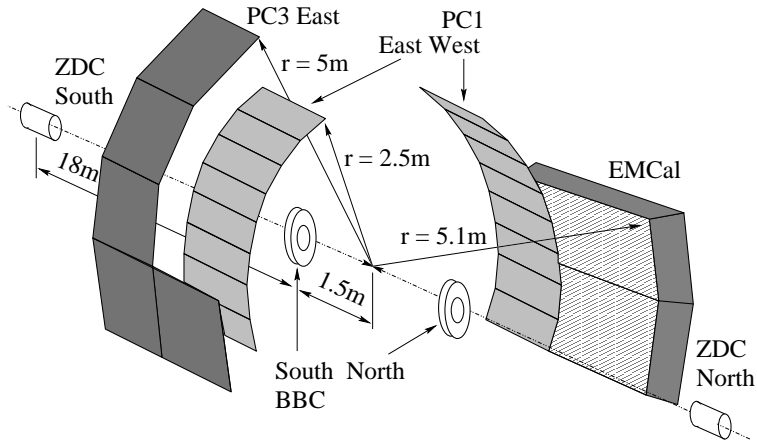,width=10cm}
\vspace{-5mm}
\caption{PHENIX detector subsystems used in this work.}
\label{fig:set-up}
\vspace{-8mm}
\end{wrapfigure}

The present analysis relies primarily on four PHENIX subsystems: the Zero Degree Calorimeters (ZDC) and the Beam-Beam Counters (BBC), used to derive the trigger and the off-line event selection, two layers of Pad Chambers (PC), called PC1 and PC3 in the East arm, used to determine the charged particle multiplicity and two sectors of the Electromagnetic Calorimeter (EMCal) in the West arm used to measure the transverse energy. These detector subsystems are shown in Fig.\ref{fig:set-up}.

\vspace{-1mm}
The ZDC are two small transverse area hadron calorimeters that measure the neutron energy within $|\eta|>6$. Besides their role as triggering devices they contribute to the centrality measurement, because their energy signal is proportional to the number of spectator neutrons~\cite{zdc}. The BBC comprise of two arrays of 64 quartz-radiator Cherenkov detectors, located at $\pm$1.44m from the center of the interaction region, covering 2$\pi$ in $\phi$ and $3.0<|\eta|<3.9$. The BBC also provide the start timing and vertex position~\cite{bbc}.

The PC provide three-dimensional coordinates along the charged-particle trajectories~\cite{pc}. Each PC layer has 8 wire chambers with cathode pad readout. PC1 and PC3 are placed at radial distances of 2.49~m and 4.98~m, respectively. The instrumented part of the EMCal is a Lead/Scintillating tile Calorimeter~\cite{rf:ieee97} which covers the pseudorapidity range $|\eta|\leq 0.38$ with an azimuthal aperture of $\Delta\phi=45^{o}$. The depth of the calorimeter corresponds to $18$ radiation lengths, equivalent to $0.85$ interaction lengths for hadrons.

The event selection consists of an interaction trigger generated by a coincidence between the north and south BBC with at least two detectors firing in each of them and a requirement of the collision vertex position, to be within $|z|\leq 17$~cm. Based on detailed simulations of the BBC, this trigger reflects [$92\pm2(syst)$]\% of the nuclear interaction cross section of 7.2 barns. More information on the trigger selection can be found in~\cite{phenix01}. The data sample used in this analysis consists of 137,784 events taken with zero magnetic field.

\vspace{-2mm}
\section{ANALYSIS}
\begin{wrapfigure}{l}{8cm}
\vspace{-9mm}
\epsfig{file=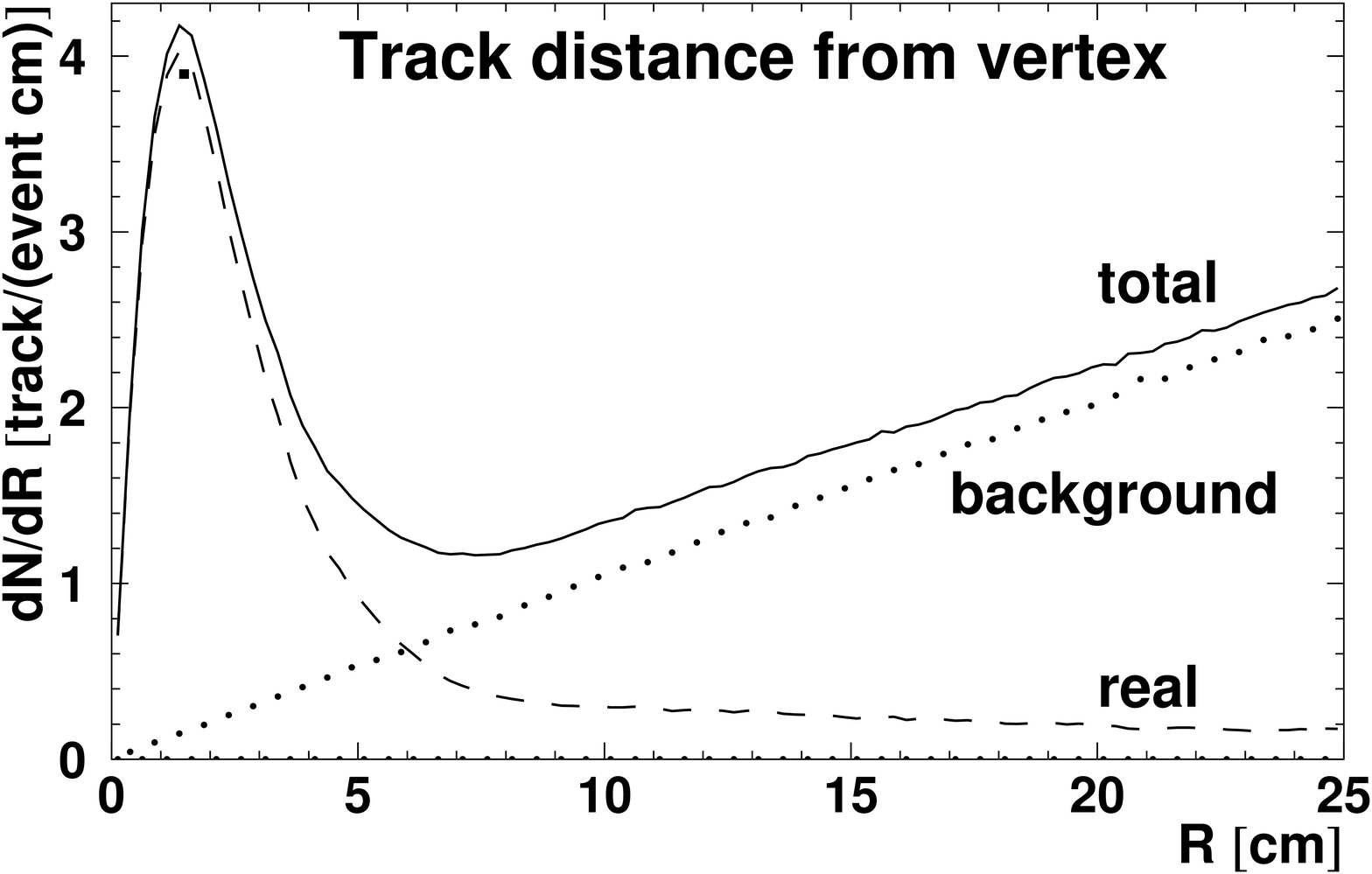,width=8cm}
\vspace{-10mm}
\caption{Track distance from vertex.}
\label{fig:rpro}
\vspace{-7mm}
\end{wrapfigure}

\vspace{-1mm}
The number of primary charged particles per event was determined on a statistical basis by correlating hits in PC1 and PC3 in the East arm, rather than by explicit track reconstruction, using the following algorithm: all hits in PC3 are combined with all hits in PC1 and the resulting lines are projected onto a plane through the beam line, symmetrically placed with respect to the tracking arm. This distribution of the distance R from the track intersection point to the event vertex is shown in Fig.\ref{fig:rpro}. Two main classes of tracks contribute to it: real tracks (peaking at low R) and tracks from the obvious combinatorial background inherent to the adopted procedure. The latter can be determined by a mixed event technique; in the present analysis, each sector in PC1 was exchanged with its neighbor and the resulting combinatorial background is shown in Fig.\ref{fig:rpro} by the dotted line. The yield of this background increases quadratically with R (leading to a linear dependence in the differential dN/dR vs. R presentation of Fig.\ref{fig:rpro}).

The R distribution of real tracks is obtained by subtracting an average background on an event-by-event basis from the total number of tracks. The sharp peak in the distribution
  at small R is due to tracks from primary particles originating at the vertex and the long tail to in-flight decay products of primary particles.

The total number of real tracks determined after background subtraction has to be corrected for several factors: corrections of 15.3\% and 1.2\% account for inactive regions and inefficiency in both PC layers, respectively. The number of tracks outside the acceptance window (25~cm) amount to 4.3\%. The track losses due to the finite double hit resolution of the PC depend on the event multiplicity. The losses are due to two effects: the direct counting of tracks (13.3\% for the 5\% most central collisions) and the subtraction of the combinatorial background by event mixing (3.6\% of the background). Finally, a 2.8\% correction takes into account the effect of charged and neutral particle decay in flight. A geometrical scaling factor of 5.82 allows to convert the corrected number of tracks to the charged particle multiplicity in one unit of pseudorapidity. The systematic errors were thoroughly studied using both the data and the guidance from detailed Monte Carlo simulations. The total systematic error amounts to 6.5\% at the highest multiplicities. See~\cite{phenix01} for more details about the analysis.

\begin{wrapfigure}{l}{8cm}
\vspace{-8mm}
\epsfig{file=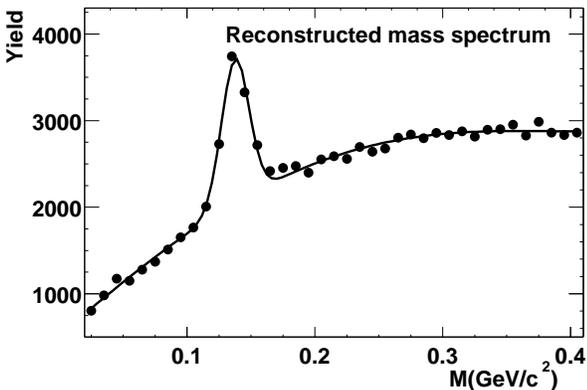,width=8cm}
\vspace{-10mm}
\centering\caption{$\pi^{o}$ peak reconstructed by EMCal.}
\label{fig:pion}
\vspace{-8mm}
\end{wrapfigure}

\vspace{2mm}
The absolute scale of the EMCal was pre-calibrated during construction by cosmic-ray muons and in-situ by monitoring the minimum ionizing peak of relativistic charged hadrons and identified high momentum electrons. The $\pi^0$ mass, reconstructed from pairs of EMCal clusters of total energy greater than 2 GeV is shown in Fig.\ref{fig:pion}, and is found to be within 1.5\% of the true value.

The transverse energy $E_T$ was computed for each event using clusters of minimum energy of 20MeV, from individual towers with deposited energy of more than 3MeV. The angle of the cluster was computed from the centroid of its energy assuming a particle originating from the event vertex. A correction factor of 1.17 converts the measured $E_T$ to the total $E_T$ in the fiducial aperture of the EMCal, and includes the response of the detector to charged and neutral particles from the event vertex and for energy in- and out-flow from the fiducial aperture. This factor was determined from a full Monte Carlo simulation of the detector, using HIJING~\cite{hijing} as event generator. A correction of 3\% accounts for disabled calorimeter towers. A scaling factor of 10.6 converts the total measured $E_T$ value into the transverse energy in one unit of pheudorapidity. The systematic error of the measurement was determined to be 4.5\% independent of multiplicity. For more details see~\cite{phenix02}.

Centrality classes were determined based on the ZDC vs. BBC response and related to the number of participating nucleons ($N_{p}$) and binary collisions ($N_{c}$) using a Glauber model~\cite{phenix01}.

\section{RESULTS}
\begin{figure}[htb]
\vspace{-6mm}
\begin{minipage}[t]{79mm}
\epsfig{file=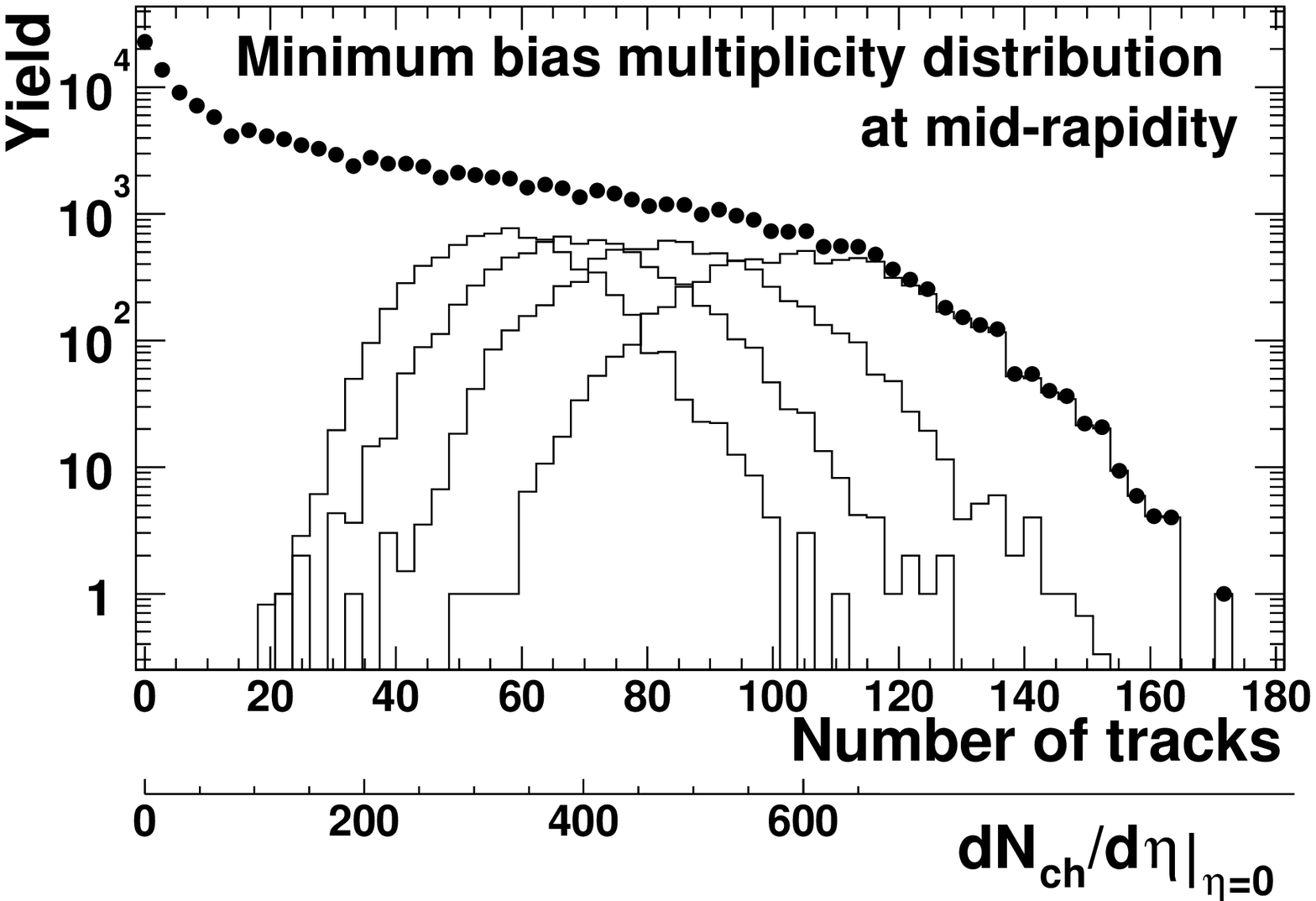,width=7.9cm}
\vspace{-17mm}
\end{minipage}
\hspace{\fill}
\begin{minipage}[t]{79mm}
\epsfig{file=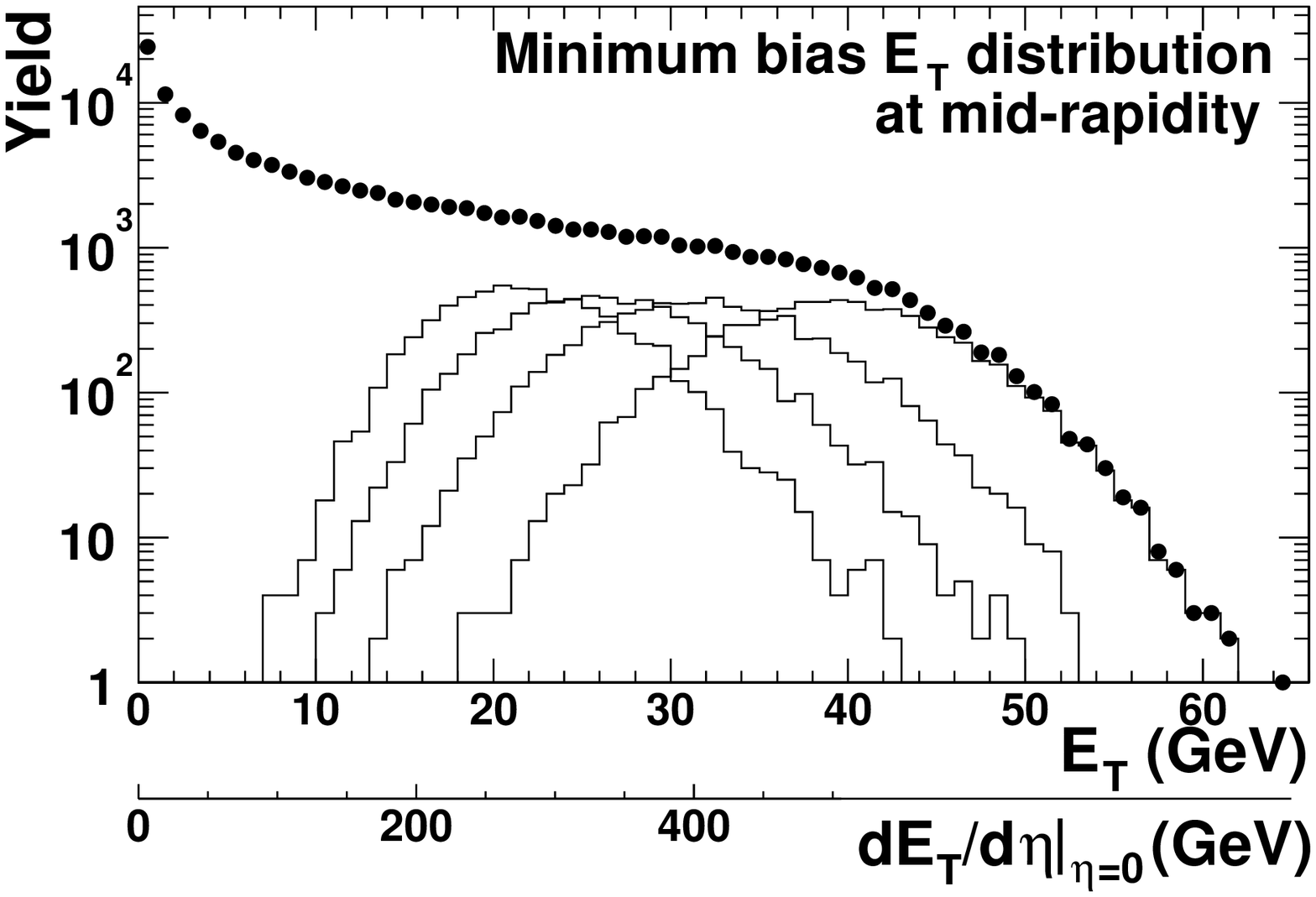,width=7.9cm}
\end{minipage}
\vspace{-17mm}
\caption{Multiplicity (left) and Transverse energy (right) densities at mid-rapidity.}
\label{fig:distr}
\vspace{-6mm}
\end{figure}

The measured distributions of $dN_{ch}/d\eta$ and $dE_{T}/d\eta$ at mid-rapidity are shown in Fig.\ref{fig:distr}. 
Both quantities show very similar behavior. The peak at low values corresponds to peripheral events, the middle region to mid-central collisions and the fall-off to the most central events. The shape of the fall-off is mainly determined by the limited detector acceptance rather than by real fluctuations. The lower axis in each plot corresponds to the measured distribution calculated for one unit of pseudorapidity. The four bell-shape curves in each plot show the distribution of the 20\% most central events in steps of 5\%.

\begin{figure}[htb]
\vspace{-6mm}
\begin{minipage}[t]{79mm}
\epsfig{file=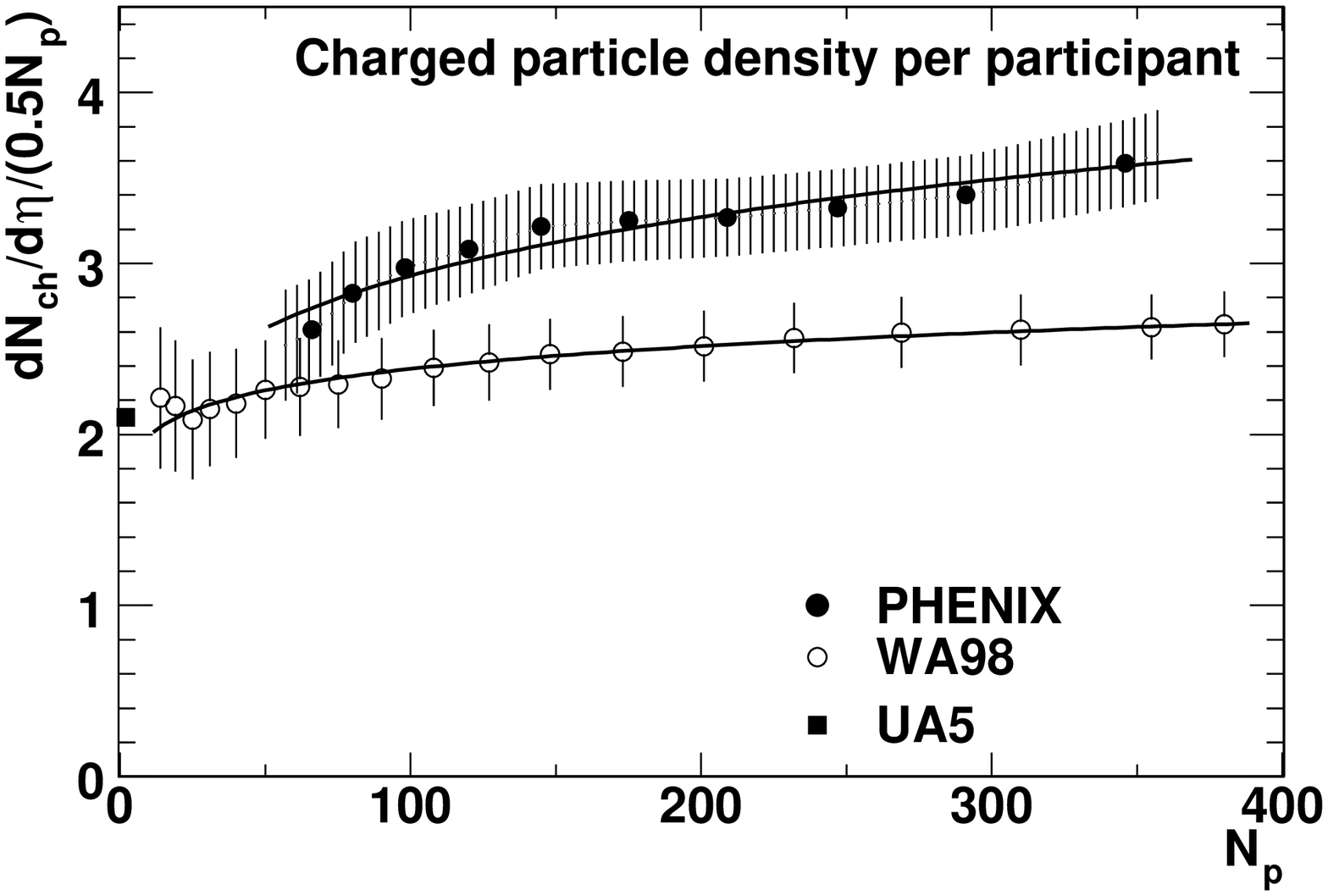,width=7.9cm}
\vspace{-17mm}
\end{minipage}
\hspace{\fill}
\begin{minipage}[t]{79mm}
\epsfig{file=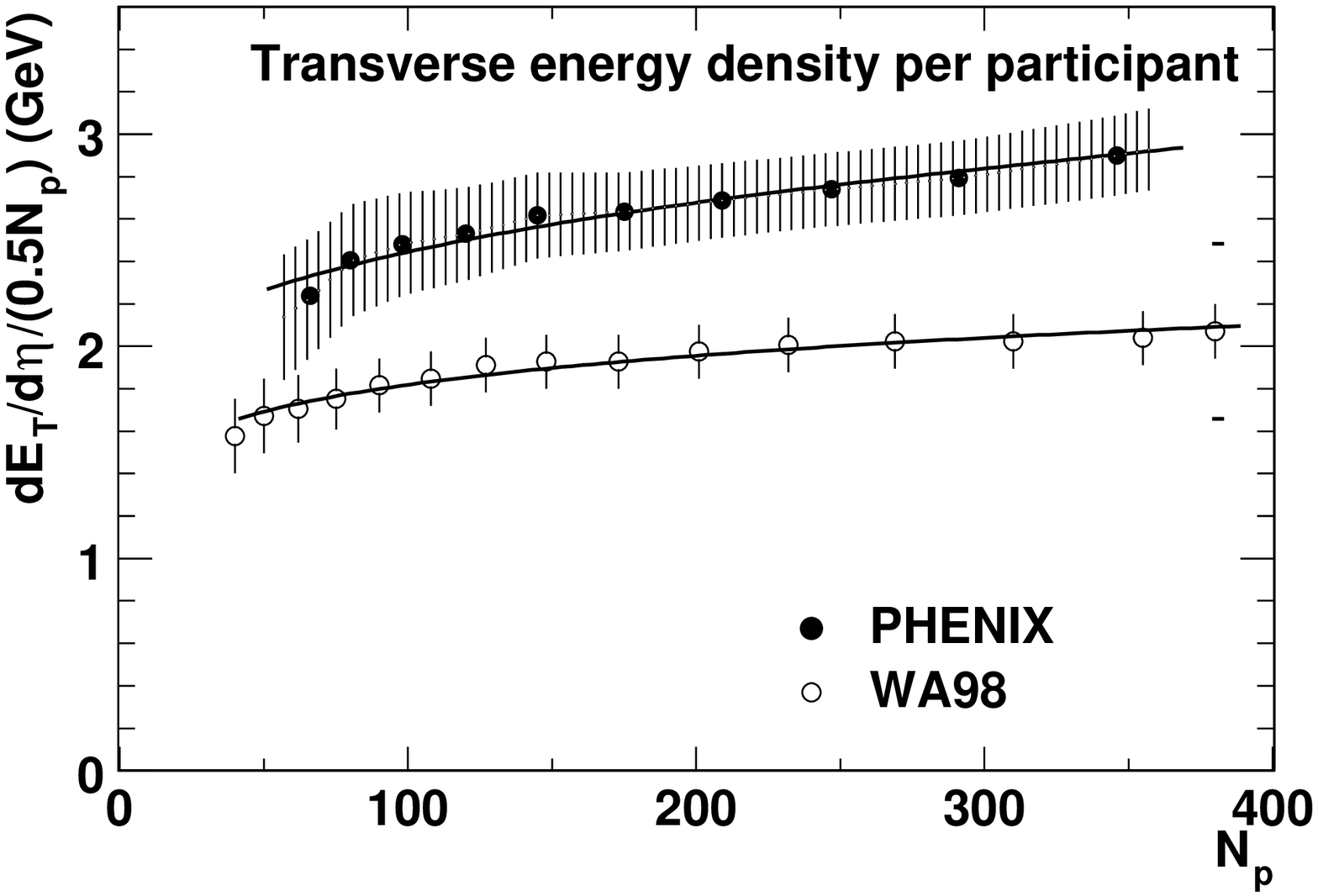,width=7.9cm}
\end{minipage}
\vspace{-17mm}
\caption{$dN_{ch}/d\eta_{|\eta=0}$ (left) and $dE_{T}/d\eta_{|\eta=0}$ (right) per pair of $N_{p}$ vs $N_{p}$.}
\label{fig:centr}
\vspace{-7mm}
\end{figure}

Fig.\ref{fig:centr} shows the multiplicity and the transverse energy per pair of $N_{p}$ vs $N_{p}$. One can see that the extrapolation to $N_{p}=2$ of the measured $dN_{ch}/d\eta_{|\eta=0}$ points is in agreement with the UA5 data for p\={p}~\cite{ua5}.
The figure also shows the results obtained by WA98~\cite{wa98} at $\sqrt{s_{_{NN}}}=17.2$~GeV. The curves represent a fit to the data points with $dX/d\eta \propto N_{p}^{\alpha}$ ($X=N_{ch}$ or $E_{T}$) as used in~\cite{wa98}. For $dN_{ch}/d\eta$ one finds $\alpha=1.16\pm0.04$ for PHENIX whereas $\alpha=1.07\pm0.04$ and $\alpha=1.05\pm0.05$ for WA98 and WA97/NA57~\cite{wa97} respectively. For $dE_{T}/d\eta$ the PHENIX value is $\alpha=1.13\pm0.05$ compared to $\alpha=1.08\pm0.06$ for WA98. One can conclude that at RHIC energies both measured quantities show a consistent increase with centrality, with an $\alpha$ value clearly larger then one, whereas at the SPS energy $\alpha$ was found to be very close to 1.

\begin{wrapfigure}{l}{8cm}
\vspace{-8mm}
\epsfig{file=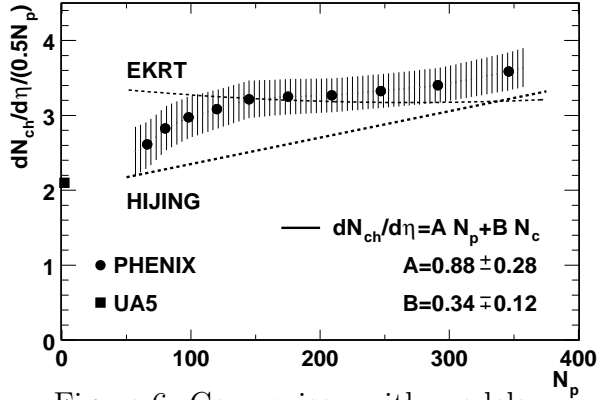,width=8cm}
\vspace{-12mm}
\centering\caption{Comparison with models.}
\label{fig:models}
\vspace{-8mm}
\end{wrapfigure}
Fig.\ref{fig:models} compares the measured $dN_{ch}/d\eta$ with the theoretical predictions of the EKRT~\cite{ekrt} saturation model and HIJING~\cite{wang-gyulassy}. The increase of $dN_{ch}/d\eta$ with $N_{p}$ is in contrast to the prediction of the EKRT model, and in qualitative agreement to HIJING, however the latter underpredicts the yield by $\sim$15\%. Following HIJING, one can assume that the increase is due to ``hard'' processes that scale with $N_{c}$ and fit the results with the parameterization $A\cdot N_{p} + B\cdot N_{c}$. One finds that the ratio $B/A$ for $dN_{ch}/d\eta$ is $0.38\pm0.19$. For the $dE_{T}/d\eta$ the ratio $B/A = 0.29\pm0.18$. The large errors are due to the negative correlation between $A$ and $B$. Therefore, in the framework of the HIJING model the ``hard'' contribution grows from $\sim$30\% for the 45-50\% centrality bin to $\sim$50\% for the most central bin (0-5\%).

\begin{figure}[htb]
\vspace{-6mm}
\begin{minipage}[t]{79mm}
\epsfig{file=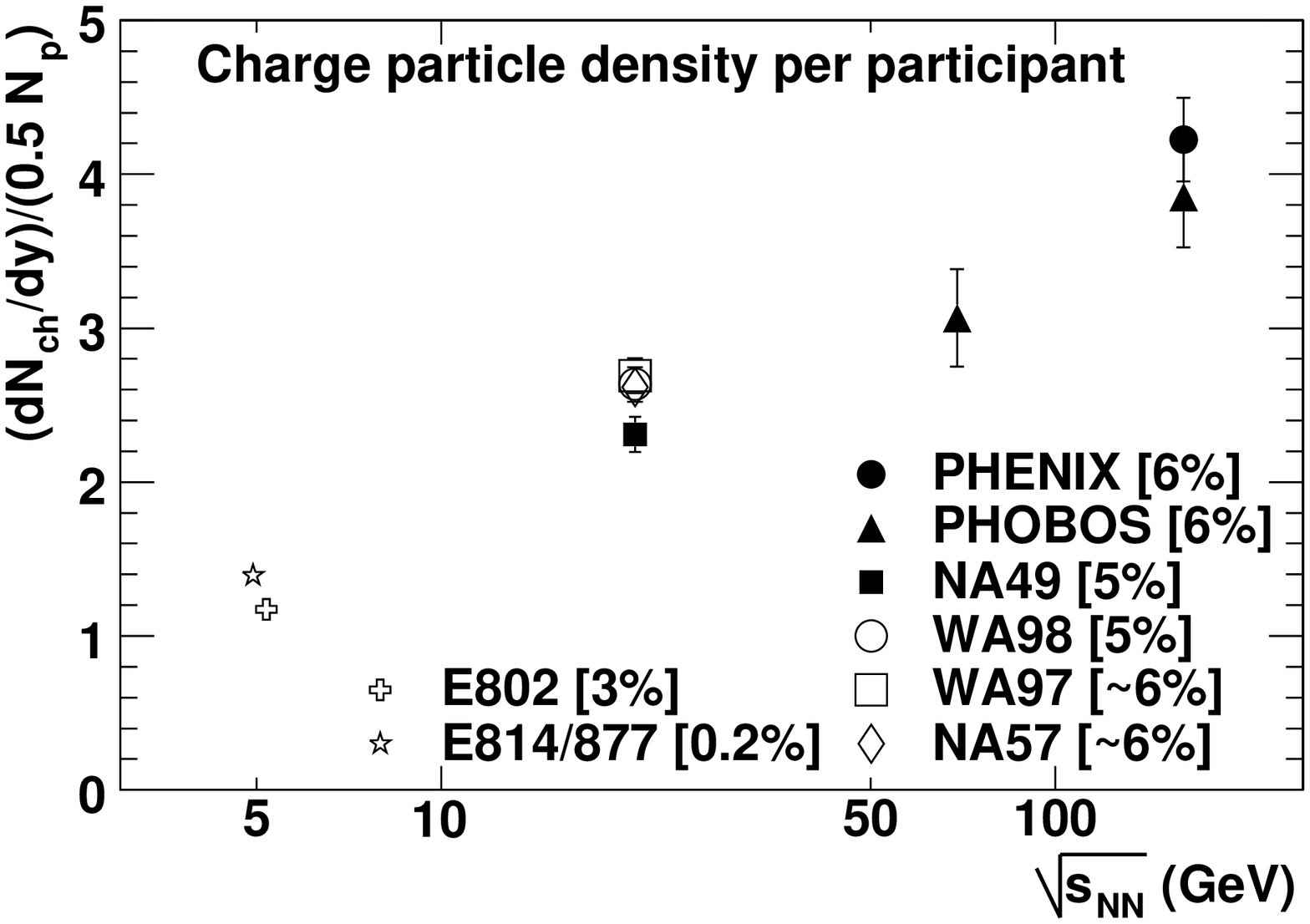,width=7.9cm}
\vspace{-17mm}
\end{minipage}
\hspace{\fill}
\begin{minipage}[t]{79mm}
\epsfig{file=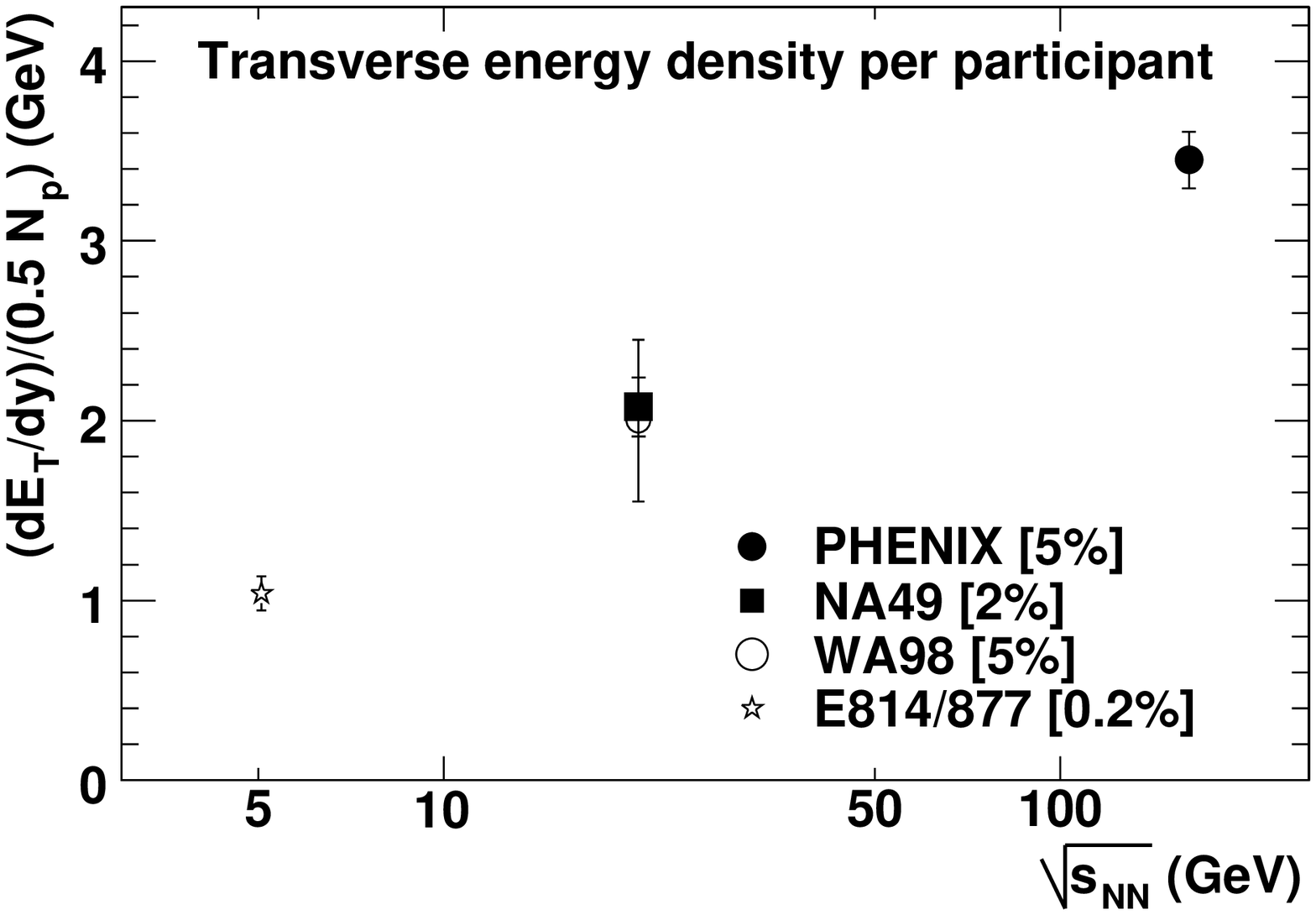,width=7.9cm}
\end{minipage}
\vspace{-15mm}
\caption{$dN_{ch}/dy_{|y=0}$ (left) and $dE_{T}/dy_{|y=0}$ (right) per pair of $N_{p}$ vs.$\sqrt{s_{_{NN}}}$ for central collisions. Data are taken from: PHENIX~\cite{phenix01,phenix02}, PHOBOS~\cite{phobos}, WA98~\cite{wa98}, WA97/NA57~\cite{wa97}, NA49~\cite{na49-1,na49-2}, E814/E887~\cite{e814-1,e814-2}, and E802~\cite{e802}.}
\label{fig:sqrts}
\vspace{-7mm}
\end{figure}

Rapidity densities per $N_{p}$ obtained in different experiments are shown in~Fig.\ref{fig:sqrts} as a function of the center of mass energy. The compiled data were measured for different centrality percentiles (values shown in brackets) in the center of mass system for PHENIX and PHOBOS or laboratory system for all other experiments. In making this figure it was assumed that $dX/dy \simeq dX/d\eta$ in the lab system and factors of 1.19 at $\sqrt{s_{_{NN}}}=130$~GeV and 1.24 at $\sqrt{s_{_{NN}}}=56$~GeV were used to account for the $\eta \rightarrow y$ transformation in the c.m. system.
The PHENIX result is in good agreement with the recent PHOBOS measurement. Comparing the mean values for RHIC and SPS shows that the particle and transverse energy production per participant increase by $\sim70$\%. The data points are consistent with $dX/dy \propto ln(\sqrt{s_{_{NN}}})$ behaviour over a broad range of energies.

\begin{figure}[htb]
\vspace{-6mm}
\begin{minipage}[t]{79mm}
\epsfig{file=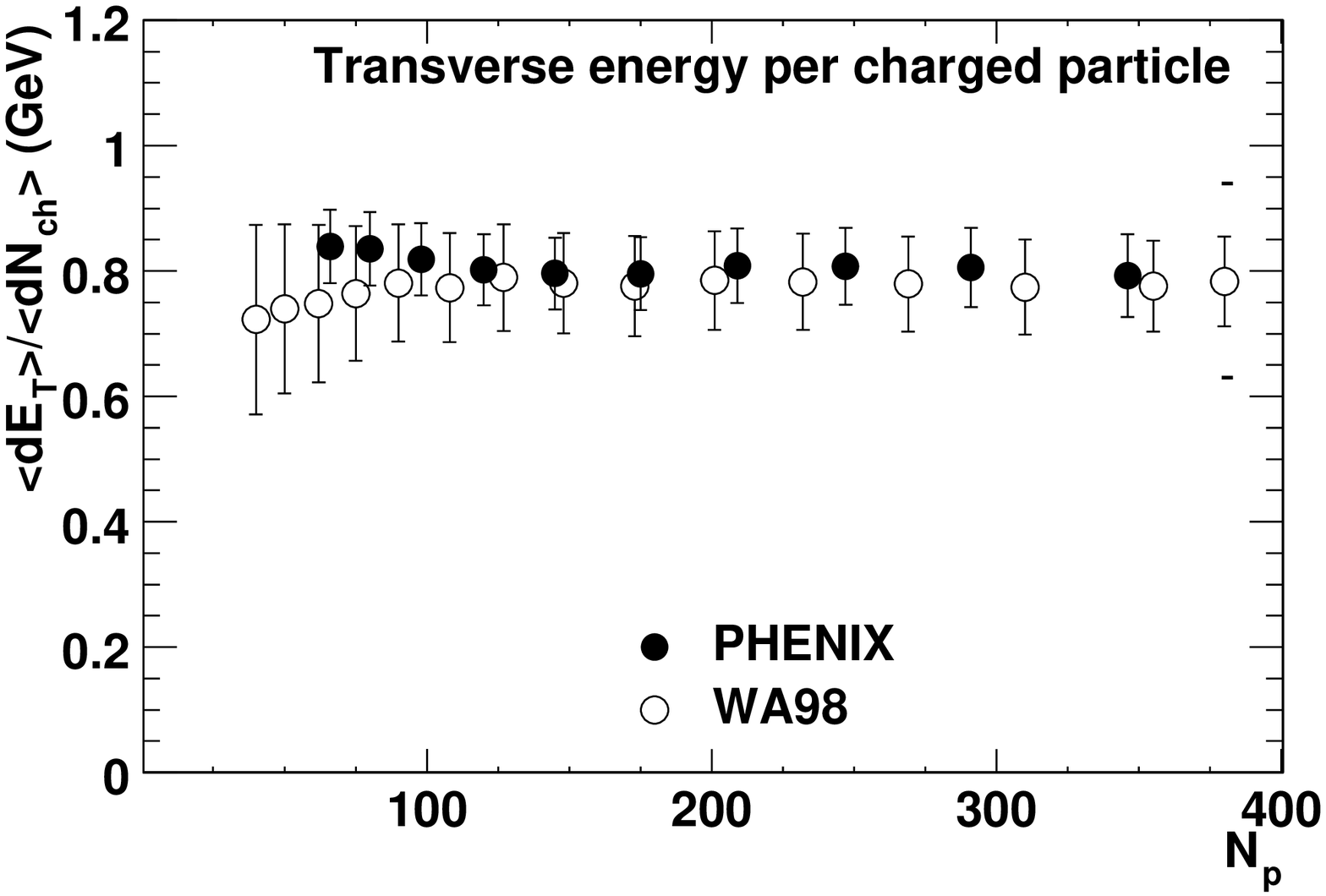,width=7.9cm}
\vspace{-17mm}
\end{minipage}
\hspace{\fill}
\begin{minipage}[t]{79mm}
\epsfig{file=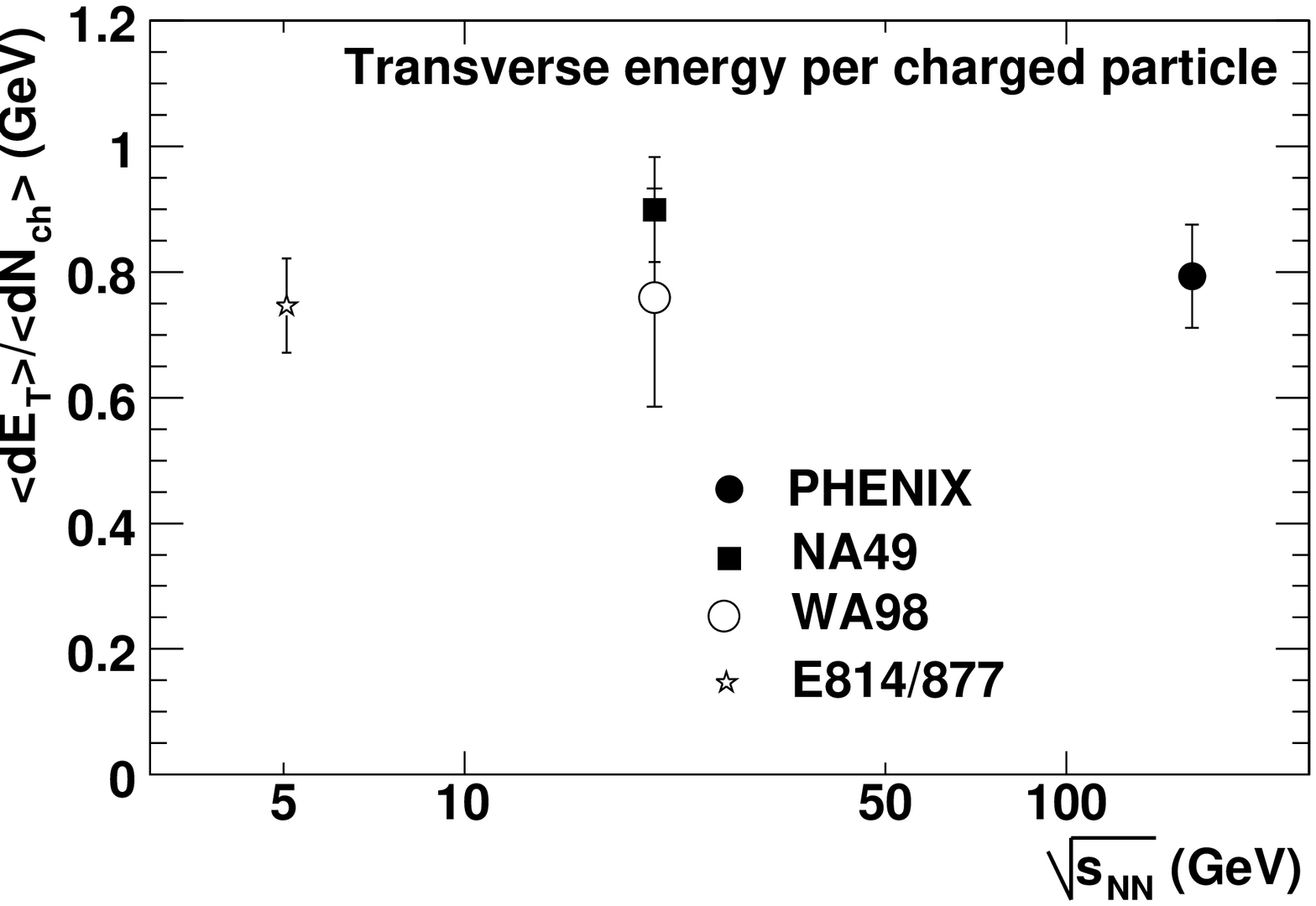,width=7.9cm}
\end{minipage}
\vspace{-17mm}
\caption{$\langle E_{T}\rangle / \langle N_{ch} \rangle$ vs. $N_{p}$ (left) and vs. $\sqrt{s_{_{NN}}}$ (right) for 0-5\% centrality bin. Data are taken from: PHENIX~\cite{phenix01,phenix02}, WA98~\cite{wa98}, NA49~\cite{na49-1,na49-2}, and E814/E887~\cite{e814-1,e814-2}.}
\label{fig:rat_sqrts}
\vspace{-7mm}
\end{figure}

Finally, Fig.\ref{fig:rat_sqrts} shows the ratio $\langle E_{T}\rangle/\langle N_{ch}\rangle$ of the transverse energy per charged particle. The ratio stays constant, equal to 0.8~GeV over a broad range of centralities, as was also observed by WA98. Moreover, the ratio, stays also constant as a function of $\sqrt{s_{_{NN}}}$, from AGS to SPS and up to RHIC energies.


\end{document}